\documentclass[preprint, 5p, twocolumn, number, sort&compress, pdftex]{elsarticle}
\usepackage[T1]{fontenc}
\usepackage[latin1]{inputenc}
\usepackage[english]{babel}
\usepackage{amssymb, amsmath}
\usepackage{bm}
\usepackage{textcomp}
\usepackage{color}
\usepackage{lmodern}
\usepackage[bookmarksnumbered=true, bookmarksopen=false, colorlinks]{hyperref}

\newcommand{\eq}{\begin{equation}} 
\newcommand{\eqe}{\end{equation}} 
\newcommand{\pa}{\partial}

\journal{Chaos, Solitons and Fractals}
\begin{document}
\title{Rayleigh--B\`enard convection in the generalized Oberbeck--Boussinesq system} 

\author[wigner,eli]{I.F.~Barna\corref{barnai}}
\cortext[barnai]{Corresponding author}
\ead{barna.imre@wigner.mta.hu}

\author[wigner,pte]{M.A.~Pocsai}

\author[elte]{S.~L\"ok\"os}

\author[sapientia]{L.~M\'aty\'as}

\address[wigner]{Wigner Research Centre for Physics of the Hungarian Academy of Sciences, Konkoly--Thege Mikl\'os \'ut 29--33, H-1121 Budapest, Hungary}
\address[eli]{ELI-HU Nonprofit Ltd., Dugonics T\'er 13, H-6720 Szeged, Hungary}
\address[pte]{University of P\'ecs, Institute of Physics, Ifj\'us\'ag \'utja 6, H-7624 P\'ecs, Hungary}
\address[elte]{E\"otv\"os Lor\'and University, Department of Atomic Physics, P\'azm\'any P\'eter s\'et\'any 1/a, H-1117 Budapest, Hungary}
\address[sapientia]{Sapientia University, Department of Bioengineering, Libert\u{a}tii sq. 1, 530104 Miercurea Ciuc, Romania}

\begin{abstract} 
The original Oberbeck--Boussinesq (OB) equations which are the coupled two dimensional Navier--Stokes(NS) and heat conduction equations have been investigated by E.N.~Lorenz half a century ago with Fourier series and opened the way to the paradigm of chaos. In our former study---Chaos, Solitons and Fractals \textbf{78,} 249 (2015)---we presented fully analytic solutions for the velocity, pressure and temperature fields with the aim of the self-similar Ansatz and gave a possible explanation of the Rayleigh--B\`enard convection cells. 
Now we generalize the Oberbeck--Boussinesq hydrodynamical system, going beyond the first order Boussinesq approximation and consider a non-linear temperature coupling. We investigate more general, power law dependent fluid viscosity or heat conduction material equations as well. Our analytic results obtained via the self-similar Ansatz may attract the interest of various fields like meteorology, oceanography or climate studies.
\end{abstract}

\begin{keyword}
Rayleigh--B\`enard convection \sep Self-similar solution \sep Oberbeck--Boussinesq approximation \sep Hydrodynamics
\end{keyword}


\maketitle

\section{Introduction}

The investigation of coupled viscous flow equations to heat conduction has a fifteen-decade long history started with 
Boussinesq \cite{bous} and Oberbeck \cite{ober} who applied it to the normal atmosphere. 
The simplest way to couple these two phenomena together is the Boussinesq approximation which is used in the field of buoyancy-driven flow (also known as natural convection). It ignores density differences except where they appear in terms multiplied by g, the acceleration due to gravity. The essence of the Boussinesq approximation is that the difference in inertia is negligible but gravity is sufficiently strong to make the specific weight appreciably different between the two fluids. Sound waves are completely impossible when the Boussinesq approximation is used since sound waves move via density variations.
Boussinesq flows are common in nature (such as katabatic winds, atmospheric fronts, oceanic circulation), industry (fume cupboard ventilation, dense gas dispersion) and the built environment (natural ventilation, central heating). Due to the approximation the calculations are easy and straightforward
to do and the interpretation are quite simple but the precision remains extremely
high. The Boussinesq approximation is applied to problems where the temperature of the fluid depends on the location, driving a flow of fluid and heat transfer. In such systems the mass, momentum and energy conservation is satisfied. In this approximation, density variations only appear when they are multiplied by g, the gravitational acceleration.
 The advantage of the approximation arises because when considering a flow of, say, warm and cold water of density $\rho_1$ and $\rho_2$ one needs only to consider a single density $\rho$: the difference $\Delta \rho = \rho_1 - \rho_2$ is negligible. The mathematics of the flow is therefore simpler because the density ratio $\rho_1/\rho_2 $, a dimensionless number, does not affect the flow; the Boussinesq approximation states that it may be assumed to be exactly one.

At the beginning of the sixties---via the stream function formalism---the Lorenz equations were derived from the Oberbeck-Boussinesq approximation to the equations describing fluid circulation in a shallow layer of fluid, heated uniformly from below and cooled uniformly from above. \cite{salz} This fluid circulation is known as Rayleigh--B\`enard convection. The fluid is assumed to circulate in two dimensions (vertical and horizontal) with periodic rectangular boundary conditions.

The partial differential equations (PDE) modeling the stream function of the system and temperature are subjected to a spectral Galerkin approximation: the hydrodynamic fields are expanded in Fourier series, which are then severely truncated to a single term for the stream function and two terms for the temperature. This reduces the model equations to a set of three coupled, nonlinear ordinary differential equations (ODE). A detailed derivation may be found, for example, in nonlinear dynamics textbooks \cite{hilb,berge}. The Lorenz system is a reduced version of a larger system studied earlier by Barry Saltzman \cite{salz}. Lorenz \cite{lorenz} evaluated the numerical solutions with computers and plotted the first strange attractor which was the advent of chaos studies. 

Investigating chaotic dynamical systems are still open up new questions and methods. 
The Lorenz equations also arise in numerous systems like simplified models for thermosyphons \cite{gor}, lasers \cite{haken}, chemical reactions \cite{port}, dynamos \cite{knob}, electric circuits \cite{cuom} and brushless DC motors \cite{hema}. 
From a technical standpoint, the Lorenz system is nonlinear, non-periodic, three-dimensional and deterministic. The Lorenz equations have been the subject of hundreds of research articles, and at least one book-length study \cite{sparrow}.

Regarding spatially extended systems the thermal convection without boundary layers is studied in Ref.~\cite{BoOr97}. The idea have been further developed and both numerical and analytical studies have been realized \cite{Calzavarini2002}, as are the longitudinal structure functions of velocity and temperature. In certain cases analytical or semi-analytical consideration are also presented for some particular situations \cite{CaDoGi2006,DoOtRe06}. One may also find a considerable number of interesting studies related to the turbulent Rayleigh--B\`{e}nard convection \cite{ChavChi2001,LoXi2010}. The dependence of viscosity on temperature have been taken into account, having an effect on dynamics \cite{SuCaGrLo03}. Convection in flows, where a given geometry may also have a role, is studied in Refs.~\cite{ScCaLo12,TeKaPeScTo2000,NiToGeTe2002}. In case of the atmosphere, the buoyancy of the moist air may lead to convection with phase changes \cite{Vaillancourt2001,Andrejczuk2004,Andrejczuk2006}. If the environmental lapse rate---the change of the temperature with altitude---is sufficiently large, then the buoyancy raises an air parcel, which can lead to cloud formation. On the other hand, a relative low environmental lapse rate brings stability \cite{SzeMa2014,Pernigotti2007,SzeMaKeGh2016}.
 
In our former study \cite{barna} we analyzed the original Boussinesq-Oberbeck PDE system with the self-similar Ansatz ending up with a non-linear coupled ODE system, however the pressure, temperature and velocity field was evaluated in analytic forms with the help of the error functions.
The main point is that instead of the well-known linear Fourier series expansion technique we applied a given Ansatz which is a completely different method to analyze a PDE or 
a PDE system. Our Ansatz is inherently capable to describe the dispersive 
and dissipative features of the solutions. Till today we could not get any remaining ODE systems (evaluated from PDE via self-similar Ansatz) which show chaotic behavior. 

\begin{figure} 
\begin{center}
\includegraphics[scale=.25]{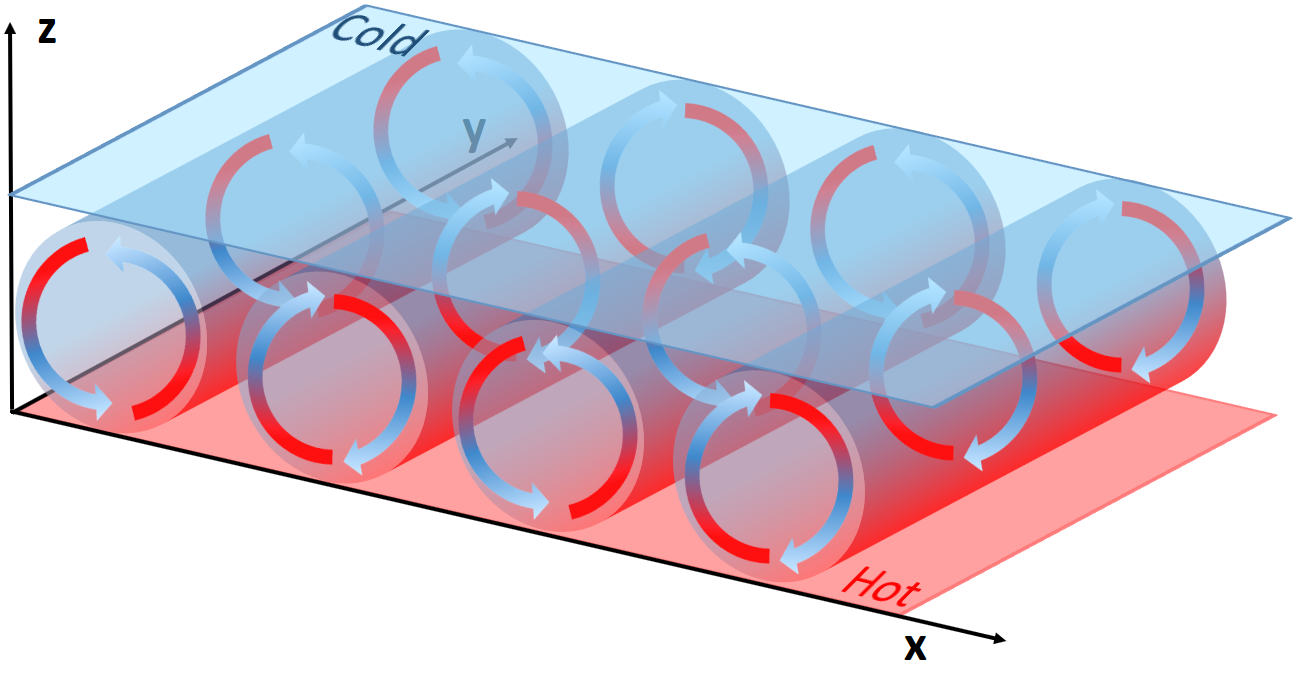}
\caption{The geometrical scheme of two dimensional convection rolls. In our modeling the x-z vertical plain is investigated as a cut of the more irregular three dimensional Rayleigh-B\`enard convection cells.}	
\label{egyes1} 
\end{center}
\end{figure}

As main result we could see the possible birth of the Rayleigh--B\`enard(RB) convection cell which is a type of natural convection, occurring in a plane horizontal layer of fluid heated from below. 
For a better understanding Fig.~\ref{egyes1} presents our recent model, which has only two spatial dimensions a horizontal and a vertical one, however it is capable to describe convection rolls, as we will see. 

Rayleigh--B\`enard convection is one of the most commonly studied convection phenomena because of its analytical and experimental accessibility. These convection patterns are the most carefully examined example of self-organizing nonlinear systems. 
Buoyancy, and hence gravity, is responsible for the appearance of convection cells. The initial movement is the upwelling of lesser density fluid from the heated bottom layer. 
 Regular pattern of cells are spontaneously organized by the upwelling. 

When, the temperature of the bottom plane is increased slightly yielding a flow of thermal energy conducted through the liquid. The system will begin to have a structure of thermal conductivity: the temperature, the pressure and density with it, will vary linearly between the bottom and top plane. A uniform linear gradient of temperature will be established. 
Once conduction is established, the microscopic random movement spontaneously becomes ordered on a macroscopic level, forming B\`enard convection cells, with a characteristic correlation length.
The rotation of the cells is a stable feature and will alternate from clock-wise to counter-clockwise horizontally, giving a nice example of spontaneous symmetry breaking. B\`enard cells are metastable which means that a small perturbation will not be able to change the rotation of the cells, but a larger one could affect the rotation, they also exhibit a form of hysteresis.

Microscopic perturbations of the initial conditions are enough to produce a non-deterministic macroscopic effect. In a repeated experiment clock-wise rotating cells may turn to counter-clockwise. Therefore there is no way to calculate the macroscopic effect of a microscopic perturbation. This inability to predict long-range conditions and sensitivity to initial-conditions are characteristics of chaotic or complex systems. Detailed physical description and exhausted technical details about the field of Rayleigh--B\`enard convection can be found in the books of \cite{ben1,ben2,ben3}. 

The connection of the self-similar Ansatz to critical phenomena, scaling, and renormalization was addressed as well. 
As far as we know all four concepts do not have a properly understood common root. 
All our studies related to two or three dimensional Navier--Stokes equations analyzed with the self-similar Ansatz can be found in the book of Campos \cite{imre_book} in Chapter 16. From our analytic velocity field with Fourier transformation additional connections to turbulence or enstropy could be evaluated as well and could be a stating point for further investigations. 

In the present study we generalize the usual Oberbeck--Boussinesq system considering more advanced material equations, like temperature dependent heat conduction coefficient, or viscosity. 
Secondly, we go beyond the simplest Boussinesq approximation which couples heat conduction to fluid dynamics as well. 
\section{Theory and Results}
To describe the problem of two dimensional heat conduction in viscous incompressible fluids one of the simplest way is the above mentioned
Oberbeck--Boussinesq model \cite{ober,salz}
\begin{equation}\label{nav} 
\begin{split}
\frac{\partial u}{\partial t} + u \frac{\partial u}{\partial x} + w \frac{\pa u}{\pa z}& + 
\frac{\pa P}{\pa x}\\ - \nu \left( \frac{\pa^2 u} {\pa x^2} + \frac{\pa^2 u} {\pa z^2} \right) = 0, \\ 
\frac{\partial w}{\partial t} + u \frac{\partial w}{\partial x} + w \frac{\pa w}{\pa z}& + 
\frac{\pa P}{\pa z} - eG T_1\\ - \nu \left( \frac{\pa^2 w} {\pa x^2} + \frac{\pa^2 w} {\pa z^2} \right) = 0,\\ 
\frac{\pa T_1}{\pa t} + u \frac{\pa T_1}{\pa x}& + w \frac{\pa T_1}{\pa z}\\ 
- \kappa \left( \frac{\pa^2 T_1 } {\pa x^2} + \frac{\pa^2 T_1} {\pa z^2} \right)& = 0,\\ 
\frac{\pa u}{\pa x} + \frac{\pa w}{\pa z}& = 0, 
\end{split}
\end{equation}
where $u,w, $ denote respectively the x and z velocity coordinates, $T_1$ is the temperature difference relative
to the average ($T_1 = T - T_{av}$) and $P$ is the scaled pressure over the density.
The free physical parameters are $\nu, e, G, \kappa $ kinematic viscosity, coefficient of volume expansion, 
acceleration of gravitation and coefficient of thermal diffusivity, respectively. 
(To avoid further misunderstanding we use $G$ for gravitation acceleration and g which is reserved for a self-similar solution.)\
The first two equations are the Navier--Stokes equations, the third one is the heat conduction equation and the last one is the continuity 
equation. All of them contain two spatial dimensions. 
We apply Cartesian coordinates and Eulerian description. 

For highly nonlinear media the temperature dependence of the density can be approximated with the following Taylor 
series 
\begin{equation}
\rho(T) = \rho_0 + \frac{\partial \rho}{\partial T }(T-T_1) + \frac{\partial^2 \rho}{\partial T^2 }(T-T_1)^2,
\end{equation}
considering the linear term only gives us the Boussinesq approximation which is presented above. 

The main goal of our forthcoming paper is to discuss physically relevant generalization of Eq.~(1) and calculate the analytic self-similar solutions. 

Our present problem can be summarized in the following PDE system:
\begin{equation}\label{nav2} 
\begin{split}
\frac{\partial u}{\partial t} + u \frac{\partial u}{\partial x} + w \frac{\pa u}{\pa z}& + 
\frac{\pa P}{\pa x}\\ - \nu\left( T_{1} \right) \left( \frac{\pa^2 u} {\pa x^2} + \frac{\pa^2 u} {\pa z^2} \right) = 0, \\ 
\frac{\partial w}{\partial t} + u \frac{\partial w}{\partial x} + w \frac{\pa w}{\pa z}& + 
\frac{\pa P}{\pa z} - eG \cdot k\left( T_{1} \right)\\ - \nu\left( T_{1} \right) \left( \frac{\pa^2 w} {\pa x^2} + \frac{\pa^2 w} {\pa z^2} \right) = 0,\\ 
\frac{\pa T_1}{\pa t} + u \frac{\pa T_1}{\pa x}& + w \frac{\pa T_1}{\pa z}\\ 
- \kappa\left( T_{1} \right) \left( \frac{\pa^2 T_1 } {\pa x^2} + \frac{\pa^2 T_1} {\pa z^2} \right)& = 0,\\ 
\frac{\pa u}{\pa x} + \frac{\pa w}{\pa z}& = 0, 
\end{split}
\end{equation}
where $ k(T_1)$ means any non-linear temperature dependence, $\nu(T_1)$ and $\kappa(T_1)$ are temperature dependent viscosities and 
heat conduction coefficients. Of course, much more complex material equations can be considered e.g~non-Newtonian fluids where 
the viscosity is velocity dependent. Analysis of such media was performed in one of our former studies \cite{imre_newtoni}.
We restrict ourselves however, ``only'' to these degrees of freedom which will open rich mathematical spectra and can give remarkable results 
for the atmosphere (air) or for oceans (water) as non-linear media. 
 
We neglect the stream function reformulation of the two dimensional flow and keep the original variables investigating the original hydrodynamical system with the Ansatz of 
\begin{equation}\label{ans}
\begin{split}
u(\eta)& = t^{-\alpha} f(\eta),\\
w(\eta)& = t^{-\delta} g(\eta),\\
P(\eta)& = t^{-\epsilon} h(\eta),\\
T_1(\eta)& = t^{-\omega} l(\eta), 
\end{split}
\end{equation}
where the new variable is $\eta = (x+z)/t^{\beta}$. 
All the five exponents $\alpha,\beta,\delta,\epsilon,\omega $ are real numbers. (Solutions with 
integer exponents are the self-similar solutions of the first kind and sometimes can be obtained from dimensional considerations 
\cite{sedov}.) The $f,g,h,l$ objects are called the shape functions of the corresponding dynamical variables. These functions should have existing first and second derivatives for the spatial coordinates and first existing derivatives for the temporal coordinate. 
 Under certain assumptions, the partial differential equations
describing the time propagation can be reduced to ordinary differential
ones which greatly simplifies the problem. This transformation
is based on the assumption that a self-similar solution
exists, i.e., every physical parameter preserves its
shape during the expansion. Self-similar solutions usually
describe the asymptotic behavior of an unbounded or a far-field
problem; the time $t$ and the space coordinate $x$ appear
only in the combination of $x/ t^{\beta}$. It means that the existence
of self-similar variables implies the lack of characteristic
lengths and times. These solutions are usually not unique and
do not take into account the initial stage of the physical
expansion process. 
By this self-similar construction we hope to find some scaling of certain physical parameters, at least for particular cases---for example at large or small times. 
This idea has certain similarities with situations where scaling properties have been used \cite{stanley}. Regarding further areas the principle of scaling occur in the study of networks \cite{barabasi} or research in connection with neural networks \cite{gri}. 
 More detailed analysis of the properties of the self-similar Ansatz is presented and discussed 
in all our former studies like \cite{barna,imre_newtoni, barna2}.

In the present study we analyze the generalization of the Oberbeck--Boussinesq approximation. We consider a non-linear temperature coupling $k(T_1) \sim T^\lambda$ in \eqref{nav2} but keeping the constant viscosity and heat conduction coefficients. In our former study the $\lambda=1$ has been chosen. In present paper $\lambda$ is a new parameter that measures the strength of the coupling between the temperature field and flow velocity field. We will see there is a non-trivial constrain between $\omega$ and $\lambda$. This relation will provide the fast decay of the strong coupling.

We also investigate the Oberbeck--Boussinesq approximation with non-constant, temperature dependent viscosity $\nu(x,z,t) \sim T(x,z,t)^{\lambda}$ case. This would have been a desired generalization because of accurate functions available for the temperature dependent viscosity for water \cite{hardy1} or for sea water \cite{hardy2} evaluated by Hardy.

The case when Oberbeck--Boussinesq approximation is generalized with constant viscosity but non-constant, temperature dependent heat conduction coefficient also can be investigated. It can be written up in the $\kappa(x,z,t) \sim T(x,z,t)^{\lambda}$ functional form. The aforementioned motivations are the same in this case as well.

We showed that a system with temperature dependent viscosity cannot be solved by the Ansatz presented in Eq.~\eqref{ans} because the relations between the exponents turn out to be contradictory. This problem does not arise in the case of the temperature dependent heat conduction but the generalization still cannot be possible because the $\lambda$ exponent turns out to be $1$ which is the linear case.

Our generalization can be summarized with the following: in Eq.~\eqref{nav2} $\nu(T_1)=\nu=$ const., $\kappa(T_1)=\kappa=$ const. and $eGk(T_1)\rightarrow bGT_1^\lambda$. It is worth to note that constant $e$ which is the coefficient of volume expansion is changed to another constant $b$ with another physical dimension.

Thanks to the free physical parameter $\lambda$, after some algebraic manipulations only some of the self-similarity exponents got fixed to the following values: $\alpha = \delta = \beta=1/2$, $\epsilon = 1$ and $\lambda \omega = 3/2$ which are called the universality relations. 
These universality relations dictate the corresponding coupled ODE system which has the following form of 

\begin{equation}\label{ode1}
\begin{split}
-\frac{f}{2} -\frac{f'\eta}{2} + ff' + gf' + h' -2\nu f'' &= 0,\\ 
-\frac{g}{2} -\frac{g'\eta}{2} + fg' + gg' + h' - bGl^{\lambda} -2\nu g'' &= 0, \\ 
 -\omega l - \frac{l'\eta}{2} + fl' +gl' -2\kappa l'' &= 0,\\
f' + g' &= 0.
\end{split}
\end{equation}
Prime means derivation in respect to $\eta$. 
From the last (continuity) equation we automatically get the $
f + g = c $ and $ f''+ g'' =0 $ conditions which are necessary in the following. 

From the third equation we get 
\begin{equation}
 2\kappa l'' + l'\left(\frac{\eta}{2}-c \right) + \frac{3l}{2\lambda} = 0,
\label{tem_shape}
\end{equation}
which is an ODE for the temperature shape function. 

The most general solutions are 
\begin{equation}\label{kum}
\begin{split}
l = c_1 M\left(\frac{3}{2\lambda},\frac{1}{2},-\frac{(2c-\eta)^2}{8\kappa} \right)& +\\ c_2 U\left(\frac{3}{2\lambda},\frac{1}{2},-\frac{(2c-\eta)^2}{8\kappa} \right),
\end{split}
\end{equation}
the Kummer's functions, for exhaustive details see the NIST Handbook\cite{NIST}. 
(We use the formal solutions obtained by Maple 12 Software [Copyright (c) Maplesoft, a division of Waterloo Inc.~1981--2008] from now on.)

The key parameter of this function is $\lambda$ which meets our physical considerations. 
$c$ is just a shifting constant and $\kappa $ scales the diffusivity of the results. 
The smaller the $\kappa$ value the sharper the main peak of the function. 
In the following we fix the $c=0$ and $\kappa = 0.5$ values. 

From the series expansion of $M(a,b,z)$ we get 
\begin{equation}
M(a,b,z) = 1 + \frac{az}{b} + \frac{(a)_2 z^2}{(b)_2 2!} +... + 
 \frac{(a)_n z^n}{(b)_n n!}, 
\end{equation}
with the $(a)_n = a(a+1)(a+2)...(a+n-1), (a) _0 =1 $ so-called 
rising factorial or Pochhammer symbol. 
If $b$ has a fix non-negative integer value (like n) then none of the solutions have poles at $b = -n$. 
For $M(a,b,z)$ if $a$ has negative integer $a = -m $ numerical value the solution is a polynomial of degree $m$ 
in $z$. In other cases, like now when $a$ is not an integer we get a convergent series for all values of $a,b$ and $z$. 
There is a connection between the two functions, 
$U$ is defined from $M$ via 
\begin{equation}
\begin{split}
U(a,b,z) = \frac{\pi}{\sin(\pi b)} 
\left[ \frac{M(a,b,z)}{\Gamma(1+a-b)\Gamma(b)} \right. \\
\left. \vphantom{U(a,b,z) = } - z^{1-b}\frac{M(1+a-b,2-b,z)}{\Gamma(a)\Gamma(2-b)} \right], 
\end{split}
\end{equation}
where $\Gamma(a)$ is the Gamma function \cite{NIST}.

From the general properties of the self-similar Ansatz we know that 
(except some pathological cases) all positive exponents mean decaying and spreading solutions in time and space. 
Our investigated system due to the dissipative NS part is so, therefore all the exponents should be positive 
which means that both $\lambda$ and $\omega $ should be positive. 
(Negative integer values of $\lambda$ define finite degree polynomials in $\eta$ which are divergent for large $\eta$ which we skip as non-physical non-dispersive solutions.)
There are three different regime available for positive $\lambda$ which describes the weakness or strengths of the 
coupling between the heat conduction and flow in the system. 
These are the followings: 
\begin{itemize}
\item{ $ 0 < \lambda <1$ where numerous oscillations occur} 
\item{$ 1 \le \lambda < 3 $ with a single aperiodic oscillation, (at $\eta \rightarrow +\infty \hspace*{2mm} l(\eta) \rightarrow 0_{-} $ )} 
\item{$ 3 \le \lambda $ where $l(\eta) > 0$. }
\end{itemize}

\begin{figure} 
\scalebox{0.4}{
\rotatebox{0}{\includegraphics{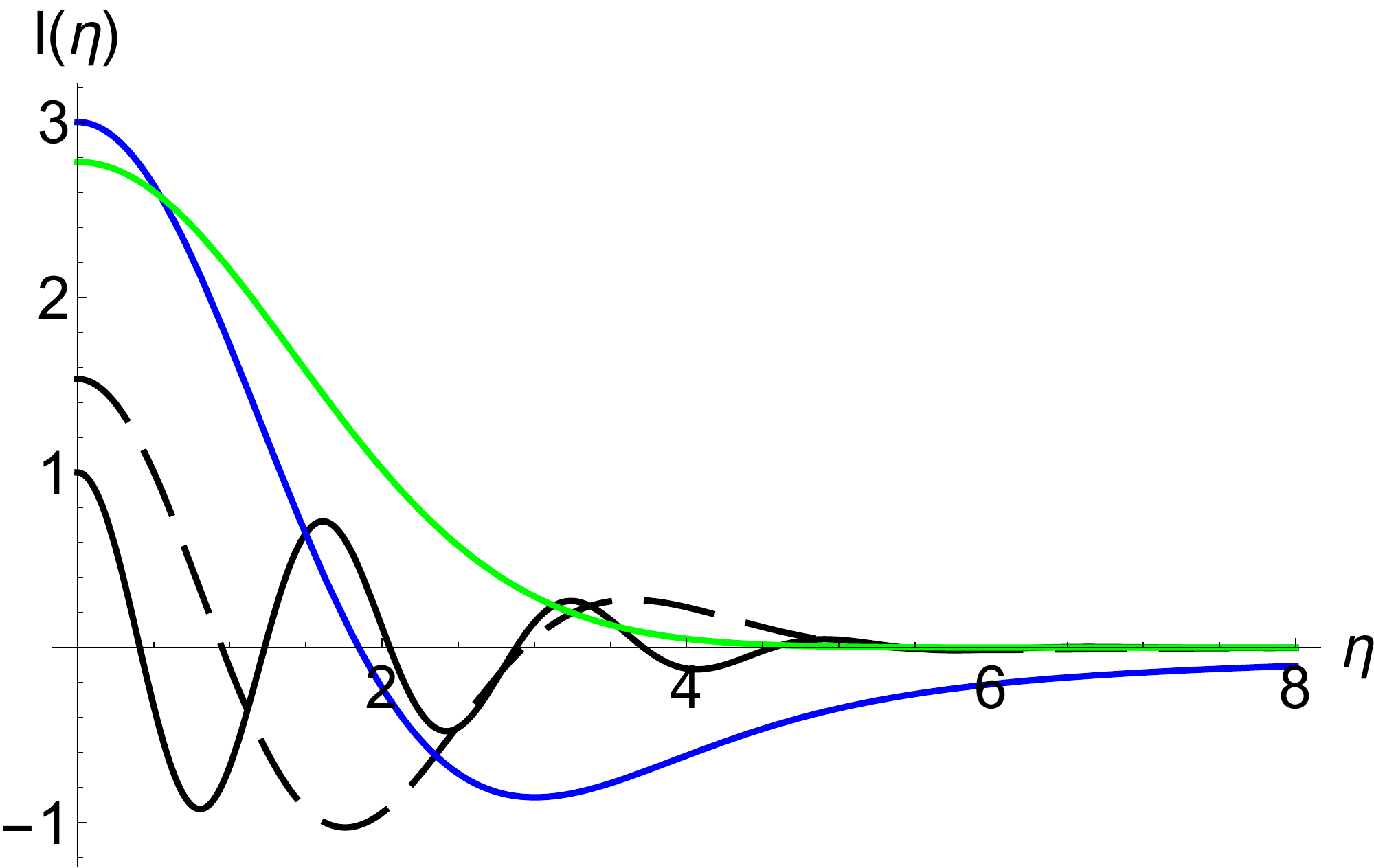}}}
\caption{The graphs of Eq.~(\ref{kum}) for $c_1=c_2=1,$ $c=0$ and $\kappa = 0.5$, only the real part was taken. 
The black solid line is for $\lambda = 0.1$, the black long dashed is for $\lambda = 0.5$ the blue solid line is for $\lambda = 1.5$,
the green solid line is for $\lambda = 3 $.}	
\label{egyes} 
\end{figure}

Figure \ref{egyes} presents such curves with different $\lambda$ values. 
The main results of our study can be clearly seen on this figure. 
With the decreasing value of $ \lambda $ (for $\lambda < 1$) the number of oscillations increase.
Which means weaker coupling 
between the temperature field and the flow velocities in the second equation of the original system. 
The term $bGT_1^{\lambda}$ in the second equation is responsible for the oscillations. 
In our former study \cite{barna} the coupling was linear, in other words the $\lambda = 1$ numerical value has been taken, 
which also means a single oscillation. 
With additional numerical investigation the number of the interceptions of the shape function versus the $\lambda$ connection can be made clear. 
Table I shows this relationship. 

\begin{table}[h!]\label{sf}
\centering
\begin{tabular}{| c | c |} \hline \hline
Value of & Nr. of zeros \\ \hline
$ 3 < \lambda $ & 0 \\ \hline
$ 1 < \lambda < 3 $ & 1 \\ \hline
$ 0.6 < \lambda < 1 $ & 2 \\ \hline
$ 0.41 < \lambda < 0.6 $ & 3 \\ \hline
$ 0.29 < \lambda < 0.41$ & 4 \\ \hline
$ 0.19 < \lambda \ 0.29$ & 6 \\ \hline
\end{tabular}
\caption{The connection between the number of interceptions and the values of $\lambda$ for $\kappa = 1/2$ and for $c_1 = c_2 = 1$, $c=0$. }
\end{table}

The argument of the temperature shape function is $\eta = (x+z)/t^{1/2}$, 
we fix the time and one of the spatial coordinates (e.g. $x$) to given values ($t_0,x_0$),
 even after this restriction there is a range of $\eta$ (or $z$) where the $l(\eta)$ function has a minimum and a maximum, 
where additional temperature and velocity fluctuation may start the Rayleigh-B\`enard convection. 
(The analysis of the velocity field clearly showed, that with fixed $t_0,x_0$ the velocity field $v_x(z)$ and $v_z(z)$ are different at the 
minimum and maximum of the temperature field, therefore the driving is present to start the convection.)
This was clearly explained in our former study \cite{barna}. The situation is very similar here, with restricted $t_0$ and $x_0$ values, 
the shape function has at least one local maximum and minimum, for $\lambda < 1$ there are two or more such oscillations, which may be the 
place of birth of parallel Rayleigh--B\`enard convection cells. It is also clear that fixing the $x$ spatial coordinate the vertical $z$ dependent convection cells are 
presented, however with a fixed $z$ spatial coordinate the horizontal $x$ convection cells are visualized. 
\begin{figure}
\begin{center}
\scalebox{0.4}{
\rotatebox{0}{\includegraphics{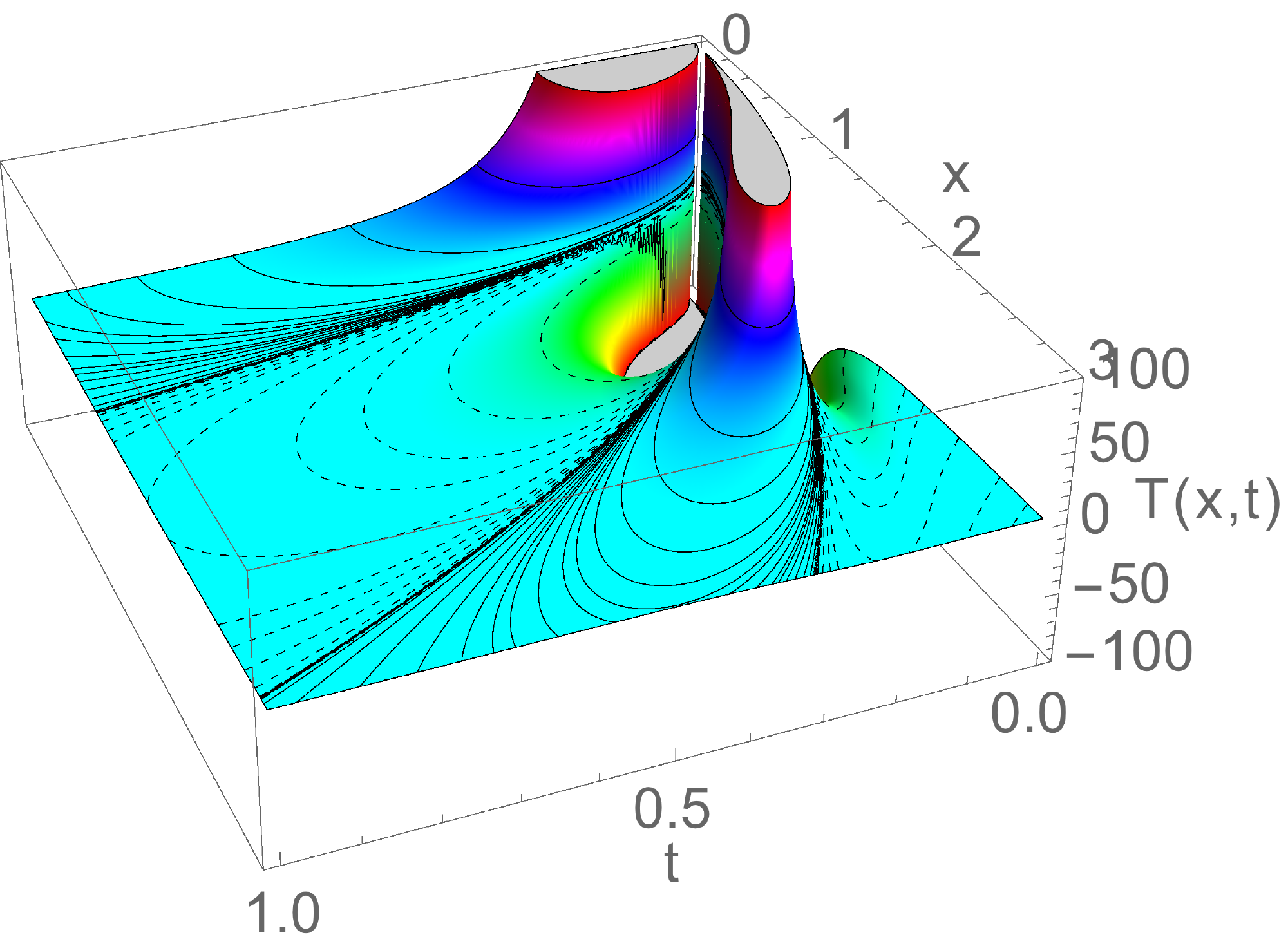}}}
\caption{The temperature field $T(x,z_0,t)$, where the real part of the solution is presented with the 
parameter set of $c_1 = c_2 =1$, $c= 0$, $\kappa = 0.5.$ with $\lambda = 0.5$. The contour lines follow a logarithmic scale. The solid and dashed lines correspond to the positive and negative values of the temperature field, respectively.}	
\end{center} 
\label{egyesl} 
\end{figure}
Figure 3 presents the three dimensional space and time dependent temperature field $T(x,z_0,t)$. 
The constrain of $\lambda \omega = 3/2$ dictates, that at low $\lambda$ values (where two or three oscillations occur)
omega has a large value ($\approx 3$) which means a quick decay in time. 

At this point for sake of completeness we mention that the solution of Eq.~(\ref{tem_shape}) for the parameters $\lambda =2$ and $c=0$ can expressed 
with the help of the modified Bessel function of the first kind $I_{\nu}(z)$ and $K_{\nu}(z)$. 
For $\lambda =3$ and $c=0$ however, the solutions are the related to the error function. 

As a second step we calculate the pressure fields as well. 
In the original OB problem, the shape function of the pressure field $h(\eta)$ is fully analytic and 
can be obtained from the temperature shape function via the following equation
\eq
h' = \frac{1}{2} \left( b G l^{\lambda} - \frac{c}{2} \right).
\label{pressure}
\eqe
Unfortunately, the general solution for an arbitrary $\lambda$ does not exist in a closed form.
Even, if we restrict the Kummer functions in Eq.~(\ref{kum}) to a pure $M$ or $U$ function, the $\lambda$ exponent makes 
it impossible to get an analytic result. 
As we mentioned above $T_1 \sim l(\eta)$ is the discrepancy from average temperature, so it can be shifted to an arbitrary level, 
it is needed because the fractional exponent value $l^{\lambda}$ has to be taken before integration of the ODE Eq.~(\ref{pressure}). 
After such a constant shift the general graph of the pressure shape function is presented on Fig. 4. 

From the first two equations one of the velocity component can be evaluated as follows 
\eq
4\nu g'' + g'(\eta -2c) + g + \frac{c}{2} + bGl^{\lambda} =0. 
\label{speed}
\eqe
Similarly to Eq.~(\ref{pressure}) there is no closed form available to the shape function of the velocity field. 
To avoid unwanted spurious cuts in the velocity field the original temperature shape function has to be shited to positive values. 
After such a transformation Figure \ref{negyes} presents a typical velocity shape function $g(\eta)$. For comparison 
we plotted it together with the original solution for $\lambda = 1$
where the solution has the form of 
\eq
g = c_1 e^{-\frac{\eta^2}{8 \nu}} erf \left( \frac{\eta}{4} \sqrt{-\frac{2}{\nu} } \right) + c_2 e^{-\frac{\eta^2}{8 \nu}} - 
c_{3}\frac{ 4eG\kappa^2 e^{-\frac{\eta^2}{8 \kappa}} }{\kappa-\nu}.
\label{geta}
\eqe 
Both parameter sets are the same. Note, that in spite of the oscillating behavior in the initial temperature shape function $l(\eta)$ 
the velocity field shape functions look very similar and smooth. The twofold integration of Eq.~(\ref{speed}) smooth out the initial temperature fluctuations.

\begin{figure}
\begin{center}
\scalebox{0.4}{
\rotatebox{0}{\includegraphics{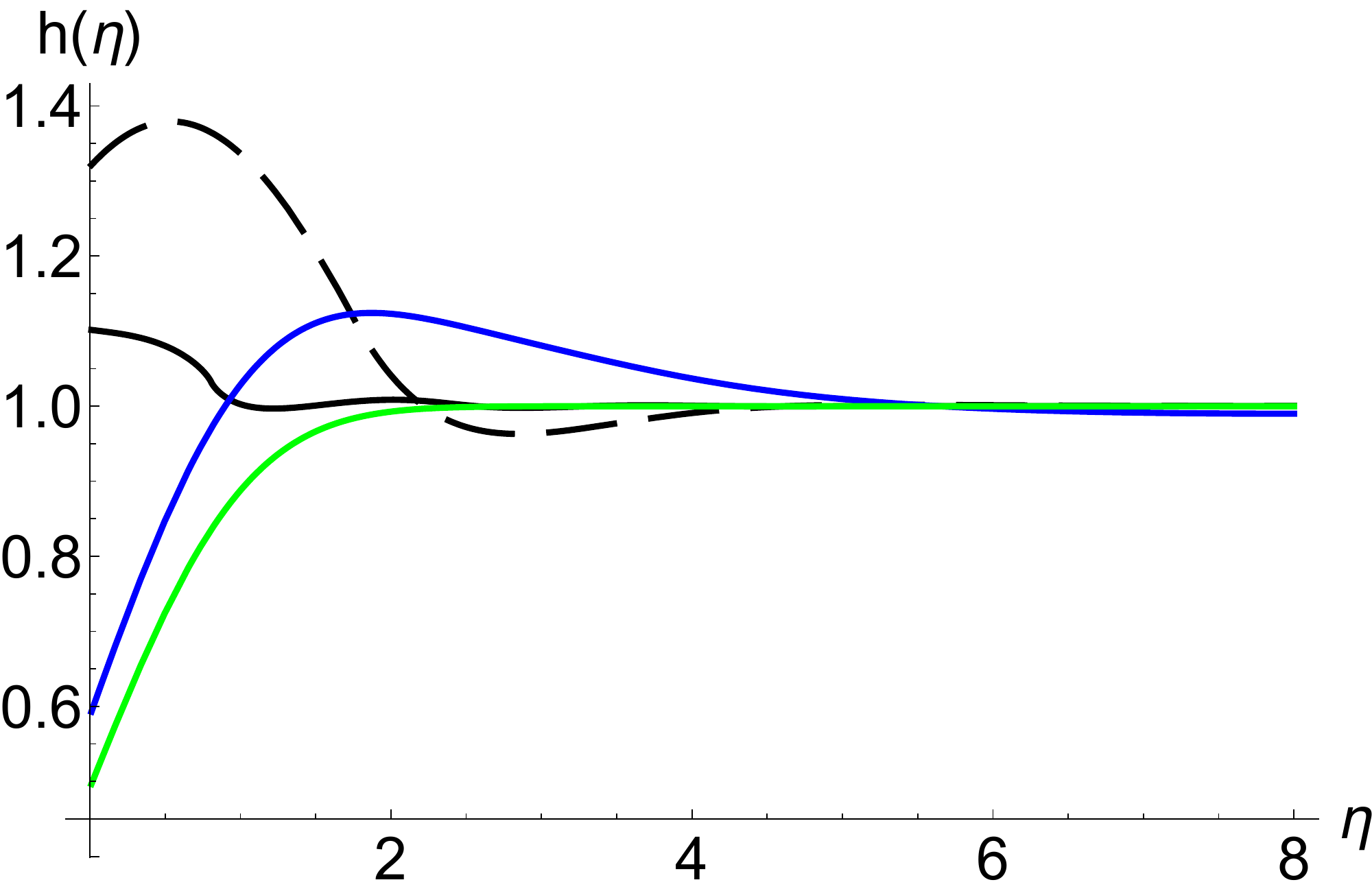}}}
\vspace*{0.4cm}
\caption{The real part of the pressure shape function $h(\eta)$ of Eq.~(\ref{pressure}) for the parameter set of $c_1 = c_2 =1$, $c= 0$, $\kappa = 0.5.$ The black solid line is for $\lambda = 0.1$, the black long dashed is for $\lambda = 0.5$ the blue solid line is for $\lambda = 1.5$, the green solid line is for $\lambda = 3. $}	
\label{harmas} 
\end{center}
\end{figure}
\begin{figure} 
\begin{center}
\scalebox{0.4}{
\rotatebox{0}{\includegraphics{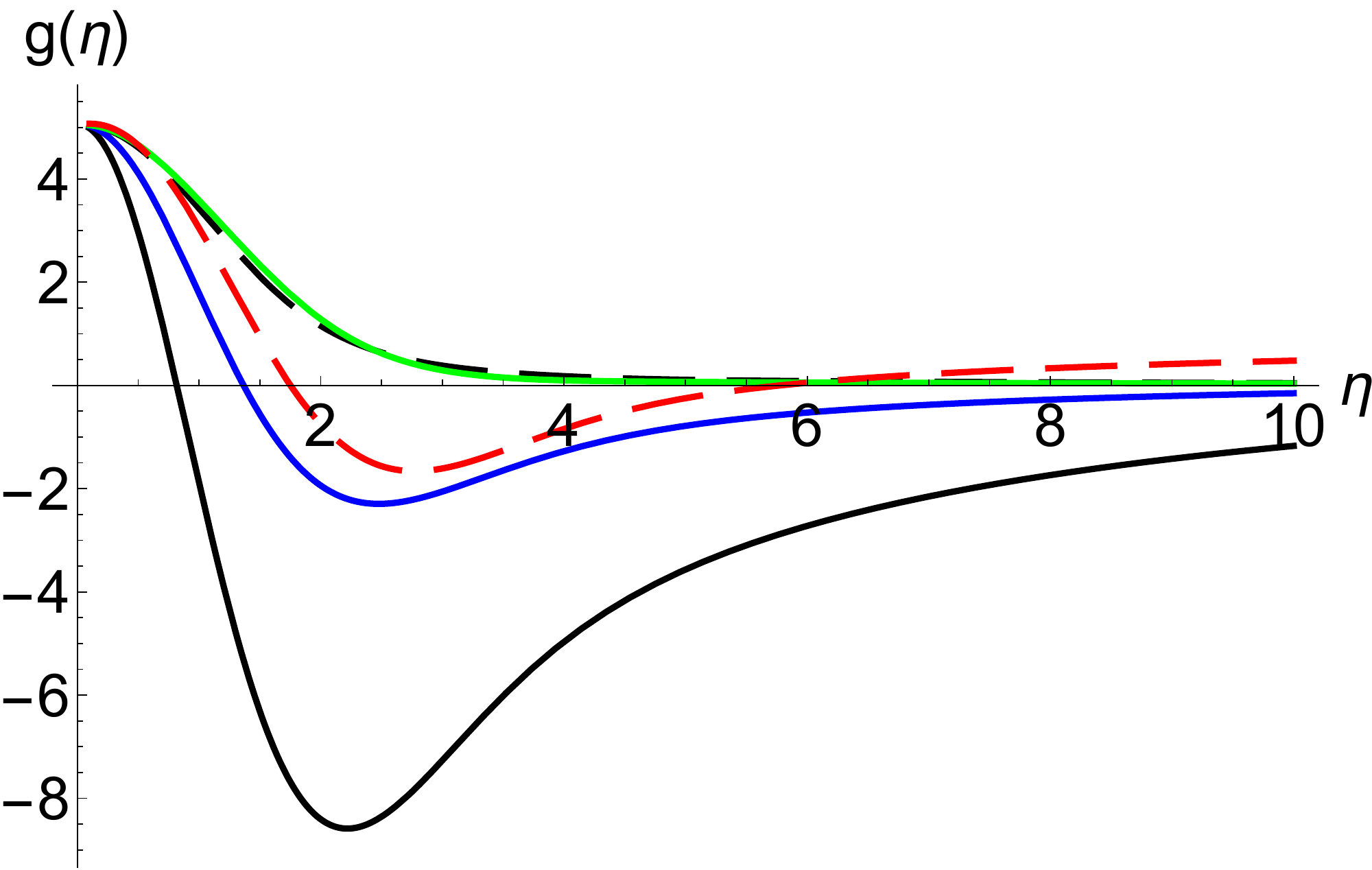}}}
\vspace*{0.4cm}
\caption{The real part of the shape function of $g(\eta)$ of \eqref{speed} which is the $z$ component of the velocity field. The black solid line is for the parameter set of $c_1 = c_2 =1$, $c= 0$, $\kappa = 0.5$, $\nu = 0.3.$, $\lambda = 0.1$, the black long dashed is for $\lambda = 0.5$ the blue solid line is for $\lambda = 1.5$, the green solid line is for $\lambda = 3 $, the red dashed line is for the former linear case of \eqref{geta} ($\lambda=1$).}
\label{negyes}
\end{center}
\end{figure}
\section{Summary and Outlook}
With reasonable generalization we investigated the classical OB equation which is the starting point of countless 
dynamical and chaotic systems. Instead of the usual Fourier truncation method we applied
the two-dimensional generalization of the self-similar Ansatz and found a coupled non-linear
ODE system which can be solved with quadrature. Our main result is that even this kind of generalization---which is beyond 
the linear Oberbeck--Boussinesq approximation---gave us an analytic temperature field which have some---not a single---oscillations. These oscillations could be the 
possible birth place of Rayleigh--B\`enard convection cells. As a second point, we may say that from the field quantities describing the system, the temperature field is the most sensitive for variations of $\lambda$.
To our best knowledge certain parts of the climate models are based on the OB
equations therefore our results might be an interesting sign to climate experts.

\end{document}